\documentstyle[12pt,aps]{revtex}

\widetext
\draft
\tighten
\begin{document}

\pagestyle{empty}

\thispagestyle{empty}

\preprint{EFUAZ FT-99-69}

\title{\normalsize{{\rm {\bf ECUACIONES DEL SEGUNDO ORDEN
EN LA REPRESENTACI\'ON $(1/2,0)\oplus (0,1/2)$ DEL
GRUPO DE LORENTZ DERIVADAS DE LOS
POSTULADOS DE RYDER}}}\thanks{Enviado a IV Reuni\'on Nacional Acad\'emica
de F\'{\i}sica y Matem\'aticas, ESFM IPN, D. F., M\'exico, 24-27 de Mayo,
1999}}

\author{Valeri V. Dvoeglazov}

\address{{\rm
Escuela de F\'{\i}sica, Universidad Aut\'onoma de Zacatecas \\
Apartado Postal C-580, Zacatecas 98068, ZAC., M\'exico\\
Email:  valeri@cantera.reduaz.mx,
URL: http://cantera.reduaz.mx/\~\,valeri/valeri.htm}
}


\maketitle

\bigskip

\begin{abstract}
\hspace*{-10mm} {\bf RESUMEN.} Utilizando los postulados de Ryder
en la representaci\'on $(1/2,0)\oplus (0,1/2)$
derivamos las ecuaciones del segundo orden.
Fueron propuestas en los sesentas y los setentas
para entender la naturaleza de los estados leptonicos
de diferentes masas.  Damos algunos discernimientos adicionales
en este problema.
\end{abstract}




\thispagestyle{empty}

\bigskip

\noindent
{\bf 1. INTRODUCCI\'ON}

\bigskip

\thispagestyle{empty}

\noindent
Una ecuaci\'on correcta para la descripci\'on adecuada de neutrinos
ha sido buscada por mucho tiempo~\cite{LY,Tok,SG,Fush1,Simon}. Este
problema, en general, est\'a conectado con el problema de l\'{\i}mite
sin masa en ecuaciones relativistas. Por ejemplo,
se sabe desde hace mucho tiempo que ``{\it  uno no puede poner simplemente
la masa igual a cero en una ecuaci\'on covariante para part\'{\i}cula
masiva, para obtener el caso correspondiente sin masa}", e.~g., ref.~[5a].

Adem\'as, en los setentas la ecuaci\'on del segundo-orden en la
reprresentaci\'on 4-dimensional del grupo $O(4,2)$  fue
propuesta por Barut {\it et al.} para resolver el problema de la
jerarqu\'{\i}a de las masas de los leptones~\cite{Bar0,Wilson,Bar}
y por Fushchich {\it et al.}, para describir varios estados del
spin en esta representaci\'on~\cite{Fush,Fush2}. Las ecuaciones (se
propusieron) dependen de dos par\'ametros. Recientemente derivamos la
ecuaci\'on de Barut y Wilson en base a los primeros
principios~\cite{DVB}.

En tercer lugar, encontramos la posibilidad de generalizaciones de las
ecuaciones en la representaci\'on $(1,0)\oplus (0,1)$ (a saber,  las
ecuaciones de Maxwell y de Weinberg, Tucker y Hammer\footnote{En general,
la \'ultima no hace completamente reducible a la anterior despu\'es de
toma el l\'{\i}mite sin masa en la ``acostumbrada" manera.}) tambi\'en en
base a incluso de dos coeficientes independientes~\cite{Dvo}, cf.
tambi\'en~\cite{Dva}.  Este nos induce a mirar al problema en
base a los primeros principios; mi investigaci\'on
se comenz\'o en~\cite{Dvo1}.

\newpage


\noindent
{\bf 2. ECUACIONES DEL SEGUNDO ORDEN}

\bigskip

\noindent
En este art\'{\i}culo nosotros  aplicamos la reformulaci\'on de
Ahluwalia~\cite{AV,DVON,DVON1} de la construcci\'on de
Majorana, McLennan y Case para neutrino~\cite{MAYOR,MLC} con
el prop\'osito de la derivaci\'on de ecuaciones pertinentes, que
revocamos  arriba.

Se usan los siguientes definiciones y postulados:

\begin{itemize}

\item

Se definen los operadores de las simetr\'{\i}as discretas como sigue:
a) el operador de la inversi\'on de espacio:
\begin{eqnarray}
S^s_{[1/2]} = \pmatrix{0&\openone\cr
\openone & 0\cr}\, ,
\end{eqnarray}
es el $4\times 4$ matriz anti-diagonal;
b) el operador de la conjugaci\'on de carga:
\begin{eqnarray}
S^c_{[1/2]} = \pmatrix{0 & i\Theta_{[1/2]}\cr
-i\Theta_{[1/2]} & 0\cr}{\cal K}\,\, , \label{cco}
\end{eqnarray}
con ${\cal K} $ siendo el operador de conjugaci\'on compleja; y
$\left (\Theta_{[j]}\right )_{h,\, h^\prime}= (-1)^{j+
h}\delta_{h^\prime,\, -h} $ siendo el  operador de Wigner.

\item
Los spinores izquierdos ($\phi_{_L}$ y $\zeta\Theta_{[j]}\phi_{_R}^\ast$)
y los derechos ($\phi_{_R}$ y $\zeta^\prime\Theta_{[j]}\phi_{_L}^\ast$)
se transforman al marco con el momento lineal $p^\mu$ (del
marco con momento cero) como sigue:
\begin{mathletters}
\begin{eqnarray}
\phi_{_R} (p^\mu)\, &=& \,\Lambda_{_R} (p^\mu \leftarrow
\overcirc{p}^\mu)\,\phi_{_R} (\overcirc{p}^\mu) \, = \, \exp (+\,{\bf J}
\cdot {\bbox \varphi}) \,\phi_{_R} (\overcirc{p}^\mu)\,\,,\\
\phi_{_L}  (p^\mu)\, &=&\, \Lambda_{_L} (p^\mu \leftarrow
\overcirc{p}^\mu)\,\phi_{_L} (\overcirc{p}^\mu) \, = \, \exp (-\,{\bf J}
\cdot {\bbox \varphi})\,\phi_{_L} (\overcirc{p}^\mu)\,\,.\label{boost0}
\end{eqnarray}
\end{mathletters}
$\Lambda_{_{R,L}} $ son las matrices de {\it boost} de Lorentz;
${\bf J}$ son las matrices de spin para $j$, e.~g., ref.~\cite{Var};
${\bbox\varphi}$ son par\'ametros del {\it boost} dado.  Si nos
restringimos por el caso de {\it bradyons} ellos se definen, {\it e.~g.},
refs.~\cite{Ryder,Dva}, por medio de:
\begin{equation}\label{boost} \cosh (\varphi) =\gamma =
\frac{1}{\sqrt{1-v^2}} = \frac{E}{m},\quad \sinh (\varphi) = v\gamma =
\frac{\vert {\bf p}\vert}{m},\quad \hat {\bbox \varphi} = {\bf n} =
\frac{{\bf p}}{\vert {\bf p}\vert}\,\, .
\end{equation}

\item
La relaci\'on de Ryder y Burgard entre spinores en el marco
con momento lineal cero\footnote{Este nombre fue introducido por D.  V.
Ahluwalia cuando consider\'o la representaci\'on $(1,0)\oplus
(0,1)$, ref.~\cite{Dva}. Si uno usa $\phi_{_R}
(\overcirc{p}^\mu)=\pm\phi_{_L} (\overcirc{p}^\mu)$, cf.
tambi\'en~\cite{Faust,Ryder}, despu\'es de aplicar las reglas de
Wigner para los {\it boosts} de los objetos con 3 componentes al momento
adquirido $p^\mu$, uno inmediatamente llega a la teor\'{\i}a de tipo de
Bargmann, Wightman y Wigner  de los campos cu\'anticos, ref.~\cite{BWW}
(cf.  tambi\'en los art\'{\i}culos viejos~\cite{Nig,Gel,Sokol} y los
recientes~\cite{ZIINO,DVOED}), en esta representaci\'on.} se
establece \begin{equation} \phi_{_L}^h (\overcirc{p}^\mu) = a
(-1)^{{1\over 2} - h} e^{i(\vartheta_1 +\vartheta_2)} \Theta_{[1/2]}
[\phi_{_L}^{-h} (\overcirc{p}^\mu)]^\ast + b e^{2i\vartheta_h}
\Xi^{-1}_{[1/2]} [\phi_{_L}^h (\overcirc{p}^\mu)]^\ast \,\, ,\label{RB}
\end{equation}
con las constantes {\it reales}
$a$ y $b$ que son arbitrarios a esta fase. $h$ es un numero cu\'antico
correspondiente a  helicidad,
\begin{eqnarray}
\Xi_{[1/2]} = \pmatrix{e^{i\phi}&0\cr
0&e^{-i\phi}\cr}\,\, ,
\end{eqnarray}
$\phi$ est\'a aqu\'{\i} el \'angulo azimuthal relacionado con
${\bf p}\rightarrow {\bf 0} $; en general, v\'ease los art\'{\i}culos
citados por la anotaci\'on.

\item
Uno puede formar o 4-spinores de Dirac:
\begin{equation}
u_h (p^\mu) =\pmatrix{\phi_{_R} (p^\mu)\cr
\phi_{_L} (p^\mu)\cr}\quad,\quad
v_h (p^\mu) = \gamma^5 u_h (p^\mu)\,\, ,
\end{equation}
o los spinores del segundo-tipo~\cite{AV}, v\'ease
tambi\'en~\cite{Sokol,DVOED}:
\begin{equation} \lambda (p^\mu) =
\pmatrix{(\zeta_\lambda \Theta_{[j]}) \phi_{_L}^\ast (p^\mu) \cr \phi_{_L}
(p^\mu)\cr}\quad,\quad \rho (p^\mu) = \pmatrix{\phi_{_R} (p^\mu) \cr
(\zeta_\rho \Theta_{[j]})^\ast \phi_{_R}^\ast (p^\mu)\cr}\,\, ,
\end{equation}
o formas  m\'as generales de 4-spinores que
dependen de los factores de la fase entre su parte izquierda
y  derecha
y sub-espacio de helicidad al cual pertenecen.
Para los espinores del segundo-tipo el autor de ref.~\cite{AV}
propuso  varias formas de los operadores del campo, e.~g.,
\begin{eqnarray}
\nu^{DL} (x^\mu) &=& \sum_\eta \int \frac{d^3 {\bf p}}{(2\pi)^3}
{1\over 2E_p} \left [ \lambda^S_\eta (p^\mu) c_\eta (p^\mu) \exp (-ip\cdot
x) +\right.\nonumber\\
&+&\left.\lambda^A_\eta (p^\mu) d_\eta^\dagger (p^\mu) \exp
(+ip\cdot x)\right ]\,\, .
\end{eqnarray}

\end{itemize}

Brevemente, el esquema para la derivaci\'on de la
ecuaci\'on  antigua
\begin{equation}
\left [i\gamma^\mu \partial_\mu  + \alpha_2 \partial^\mu
\partial_\mu  -\kappa \right ] \phi (x^\mu)
= 0\,\,   \label{Barut}
\end{equation}
es la siguiente. Primero, aplica la relaci\'on generalizada de
Ryder y Burgard (v\'ease arriba, Eq. (\ref{RB})) y el esquema
normal para la derivaci\'on de las ecuaciones relativistas de
la onda~\cite[nota de pie \# 1]{AV}. Despu\'es, forma  los
 4-spinores de Dirac; sus partes derechas e izquierdas se conectan
como sigue:
\begin{mathletters} \begin{eqnarray}
\phi_{_L}^\uparrow (p^\mu) &=& - \Theta_{[1/2]} [\phi_{_R}^{\downarrow}
(p^\mu)]^\ast \quad,\quad \phi_{_L}^\downarrow (p^\mu) = + \Theta_{[1/2]}
[\phi_{_R}^{\uparrow}(p^\mu)]^\ast \,\, ,\label{1a}\\
\phi_{_R}^\uparrow (p^\mu) &=& -
\Theta_{[1/2]} [\phi_{_L}^{\downarrow} (p^\mu)]^\ast \quad,\quad
\phi_{_R}^\downarrow (p^\mu) = + \Theta_{[1/2]}
[\phi_{_L}^{\uparrow} (p^\mu)]^\ast \label{1b}\,\,,
\end{eqnarray}
\end{mathletters}
para  obtener
\begin{equation}
\left [a \,{i\gamma^\mu \partial_\mu \over m}
+b\, {\cal C} {\cal K} - \openone\right ] \Psi (x^\mu) = 0\,\, ,\label{de}
\end{equation}
en el espacio de
coordenadas.  Transferir a la representaci\'on de Majorana con la
matriz unitaria
\begin{equation} U ={1\over
2}\pmatrix{\openone -i\Theta_{[1/2]} & \openone +i\Theta_{[1/2]}\cr
-\openone -i\Theta_{[1/2]} & \openone -i\Theta_{[1/2]}\cr}\quad,\quad
U^\dagger = {1\over 2}\pmatrix{\openone -i\Theta_{[1/2]} & -\openone
-i\Theta_{[1/2]}\cr \openone + i\Theta_{[1/2]} & \openone
-i\Theta_{[1/2]}\cr}\,\,.\label{maj}
\end{equation}
Finalmente, uno obtiene el conjunto
\begin{mathletters}
\begin{eqnarray}
\left [ a {i\gamma^\mu \partial_\mu \over m} -\openone \right ] \phi - b
\,\chi &=& 0 \,\,,\\
\left [ a {i\gamma^\mu \partial_\mu \over m} -\openone \right ]\chi  - b
\,\phi &=& 0\,\,
\end{eqnarray}
\end{mathletters}
para $\phi (x^\mu)=\Psi_1+\Psi_2$ o $\chi (x^\mu)=\Psi_1-\Psi_2$
(donde $\Psi^{^{MR}} (x^\mu)=\Psi_1+ i\Psi_2$).
Con la identificaci\'on $a/ 2m\rightarrow\alpha_2$ y
$m (1-b^2)/ 2a\rightarrow\kappa$ el conjunto de ecuaciones lleva a la
ecuaci\'on del segundo-orden del tipo de Barut.

En la base de definiciones anteriores
usando las reglas normales~\cite[nota a pie \# 1]{AV} uno puede
derivar:

\begin{itemize}

\item

En el caso $\vartheta_1= 0$, $\vartheta_2=\pi$ se obtienen las
ecuaciones siguientes para $\phi_{_L} (p^\mu) $ y
$\chi_{_R}=\zeta_\lambda\Theta_{[1/2]}\phi_{_L}^\ast
(p^\mu)$:\footnote{Los factores de la fase $\zeta$ son
definidos por varias restricciones impuestos a los  4-spinores
(o a los operadores correspondientes), e.~g., la condici\'on de
auto/contr-auto conjugaci\'on de carga nos da $\zeta_\lambda^{S,A}=\pm i$.
Pero, uno todav\'{\i}a debe notar que estos factores de la fase tambi\'en
dependen del factor de la fase en la definici\'on del operador de la
conjugaci\'on de la carga (\ref{cco}). El ``t\'ermino de la masa" de
ecuaciones dinamicas resultantes tambi\'en estar\'{\i}a diferente.}
\begin{mathletters}
\begin{eqnarray}
\phi_{_L}^h (p^\mu ) &=& \Lambda_{_L} (p^\mu \leftarrow \overcirc{p}^\mu)
\phi_{_L}^h (\overcirc{p}^\mu)
= {a \over \zeta_\lambda}
(-1)^{{1\over 2} +h} \Lambda_{_L} (p^\mu \leftarrow \overcirc{p}^\mu)
\Lambda_{_R}^{-1} (p^\mu \leftarrow \overcirc{p}^\mu) \chi_{_R}^h (p^\mu)+
\nonumber\\
&+&{b\over \zeta_\lambda} \Lambda_{_L} (p^\mu \leftarrow \overcirc{p}^\mu)
\Xi^{-1}_{[1/2]} \Theta^{-1}_{[1/2]} \Lambda_{_R}^{-1} (p^\mu \leftarrow
\overcirc{p}^\mu) \chi_{_R}^{-h} (p^\mu)\,\,,\\
\chi_{_R}^{-h} (p^\mu) &=& \Lambda_{_R}
(p^\mu \leftarrow \overcirc{p}^\mu) \chi_{_R}^{-h}
(\overcirc{p}^\mu) =
a \zeta_\lambda (-1)^{{1\over 2}
-h} \Lambda_{_R} (p^\mu \leftarrow \overcirc{p}^\mu) \Lambda_{_L}^{-1}
(p^\mu \leftarrow \overcirc{p}^\mu) \phi_{_L}^{-h} (p^\mu) +\nonumber\\
&+& b\zeta_\lambda \Lambda_{_R} (p^\mu \leftarrow \overcirc{p}^\mu)
\Theta_{[1/2]} \Xi_{[1/2]} \Lambda_{_L}^{-1} (p^\mu \leftarrow
\overcirc{p}^\mu) \phi_{_L}^{h} (p^\mu)\,\,.
\end{eqnarray}
\end{mathletters}

Por consecuencia, las ecuaciones para los 4-spinores $\lambda^{S,A}_\eta
(p^\mu)$ toman las formas:\footnote{$\eta$ es el n\'umero cu\'antico
de la helicidad quiral introducido en ref.~\cite{AV}.}
\begin{mathletters}
\begin{eqnarray}
ia
{\widehat p \over m} \lambda^S_\uparrow (p^\mu) - (b{\cal C}{\cal K}
-\openone) \lambda^S_\downarrow (p^\mu) &=& 0\,\, ,
\label{m1}\\
ia {\widehat p \over m} \lambda^S_\downarrow (p^\mu) + (b
{\cal C}{\cal K} -\openone) \lambda^S_\uparrow (p^\mu) &=& 0\,\, ,
\label{m2}\\
ia {\widehat p \over m} \lambda^A_\uparrow (p^\mu) - (b
{\cal C}{\cal K} +\openone) \lambda^A_\downarrow (p^\mu) &=& 0\,\, ,
\label{m3}\\
ia {\widehat p \over m} \lambda^A_\downarrow (p^\mu) + (b {\cal
C}{\cal K} +\openone) \lambda^A_\uparrow (p^\mu) &=& 0\,\,
\label{m4},
\end{eqnarray} \end{mathletters}
$a=\pm (b-1) $ si queremos tener $p_0^ 2-{\bf p}^ 2=
m^ 2$ para part\'{\i}culas masivas.

\item
Podemos escribir varias formas de ecuaciones en la representaci\'on de
coordenadas dependiendo de las relaciones entre operadores de creaci\'on y
operadores de la aniquilaci\'on.  Por ejemplo, la primera generalizaci\'on
en el espacio de coordenadas  lee (con tal de que implicamos $d_\uparrow
(p^\mu)=+ ic_\downarrow (p^\mu)$ y $d_\downarrow (p^\mu)=- ic_\uparrow
(p^\mu) $; el operador ${\cal K} $
act\'ua a  $q-$ numeros como conjugaci\'on hermitiana)
\begin{equation}
\left [ ia {\gamma^\mu \partial_\mu \over m} - (b-\openone) \gamma^5 {\cal
C} {\cal K} \right ] \Psi (x^\mu) = 0\,\, .
\end{equation}

Transferiendo en la representaci\'on de Majorana  uno
obtiene dos ecuaciones reales:\footnote{Parece  que se puede
llevar a cabo este procedimiento para cualquier spin, cf.~\cite{Dvo2}.}
\begin{mathletters} \begin{eqnarray}
ia {\gamma^\mu \partial_\mu \over m} \Psi_1 (x^\mu) -i (b-\openone)
\gamma^5 \Psi_2 (x^\mu) &=& 0\,\, ,\\
ia {\gamma^\mu \partial_\mu \over
m} \Psi_2 (x^\mu) -i (b-\openone)\gamma^5 \Psi_1 (x^\mu) &=& 0\,\, .
\end{eqnarray} \end{mathletters}
para partes reales e imaginarias de la funci\'on de campo $\Psi^{^{MR}}
(x^\mu)=\Psi_1 (x^\mu)+ i\Psi_2 (x^\mu) $. En el
caso de $a= 1-b$ y considerando la funci\'on del campo $\phi=\Psi_1+\Psi_2$
venimos a la ecuaci\'on de Sokolik  para spinores del {\ segundo
tipo}~\cite[Eq. (8)]{Sokol} y ref.~\cite[Eqs. (14,18)]{DVOED}.
Entonces, venimos a la ecuaci\'on del segundo-orden en la
representaci\'on de coordenadas para part\'{\i}culas masivas
\begin{equation} \left [ a^2
{\partial_\mu \partial^\mu \over m^2} +(b-1)^2 \right ] \cases{\Psi_1
(x^\mu) &\cr \Psi_2 (x^\mu) &\cr} = 0\,\,.\label{kg} \end{equation}
Por supuesto, se
reduce a la  ecuaci\'on de Klein y Gordon.  En general, all\'{\i}
existir\'{\i}a fraccionamiento de las masas entre varios estados $CP-$
conjugados.  Volveremos a responder esta pregunta en otros papeles.

\item
Uno puede encontrar la relaci\'on entre operadores de creaci\'on y
aniquilaci\'on para otra ecuaci\'on ($\beta_1,\,\beta_2\in \Re e$)
m\'as general
\begin{equation}
\left [ ia {\gamma^\mu \partial_\mu \over m} - e^{i\alpha_1}\beta_1
\gamma^5 {\cal C} {\cal K}  + e^{i\alpha_2} \beta_2 \right ] \Psi (x^\mu)
= 0\,\, ,\label{neq1}
\end{equation}
que estar\'{\i}a consistente con las ecuaciones
(\ref{m1}-\ref{m4}).\footnote{Como uno puede esperar de esta
consideraci\'on, la ecuaci\'on
(\ref{neq1}) estar\'{\i}a recordatoria a los trabajos del los 60s,
refs.~\cite{SG,Fush1,Simon,Rasp}.} Aqu\'{\i} est\'an:
\begin{mathletters}
\begin{eqnarray}
(b-1) c_\uparrow &=& ie^{i\alpha_1}\beta_1 d_\downarrow -ie^{i\alpha_2}
\beta_2 c_\downarrow\,\, ,\\
(b-1) c_\downarrow &=& -ie^{i\alpha_1}\beta_1 d_\uparrow +ie^{i\alpha_2}
\beta_2 c_\uparrow\,\, ,\\
(b-1) d_\uparrow^\dagger &=& -ie^{i\alpha_1}\beta_1 c_\downarrow^\dagger
-ie^{i\alpha_2} \beta_2 d_\downarrow^\dagger\,\, ,\\
(b-1) d_\downarrow^\dagger &=& ie^{i\alpha_1}\beta_1 c_\uparrow^\dagger
+ie^{i\alpha_2} \beta_2 d_\uparrow^\dagger\,\, .
\end{eqnarray}
\end{mathletters}
La condici\'on de la compatibilidad
asegura que $\alpha_2= 0,\pi$ y $\beta_1^2+\beta_2^2= (b-1)^2$.
Asumimos que estos dos operadores de la aniquilaci\'on est\'an
independientes linealmente. Si $\beta_1= 0$  recuperamos la
ecuaci\'on de Dirac pero con restricciones adicionales puestos a los
operadores de creaci\'on y operadores de la aniquilaci\'on, $c_\uparrow
=\mp ic_\downarrow$ y $d_\uparrow=\pm i d_\downarrow$.
El factor de la fase $\alpha_1$ queda indeterminado.

En la representaci\'on de Majorana el juego resultante de las ecuaciones
reales es
\begin{mathletters}
\begin{eqnarray}
&&\left [ia {\gamma^\mu \partial_\mu \over m} +i\beta_1 \sin\alpha_1
\gamma^5  +\beta_2 \right ] \Psi_1 -i\beta_1 \cos\alpha_1 \gamma^5 \Psi_2
= 0\,\, ,\\
&&\left [ia {\gamma^\mu \partial_\mu \over m} -i\beta_1 \sin\alpha_1
\gamma^5  +\beta_2 \right ] \Psi_2 -i\beta_1 \cos\alpha_1 \gamma^5 \Psi_1
= 0\,\, .
\end{eqnarray} \end{mathletters}
Por ejemplo en el caso $\alpha_1={\pi\over 2}$  obtenemos
\begin{mathletters}
\begin{eqnarray}
\left [ ia {\gamma^\mu \partial_\mu \over m} +i \beta_1 \gamma^5 +\beta_2
\right ] \Psi_1 &=& 0\,\, ,\label{gf1}\\
\left [ ia {\gamma^\mu \partial_\mu \over m} -i \beta_1 \gamma^5 +\beta_2
\right ] \Psi_2 &=& 0\,\, \label{gf2}.
\end{eqnarray}
\end{mathletters}
Pero, en todos los casos uno puede recuperar la ecuaci\'on de
Klein y Gordon  para ambas partes, reales e imaginarias, de la
funci\'on del campo, Eq.  (\ref{kg}).  No est\'a claro todav\'{\i}a si
las estructuras discutidas recientemente en ref.~\cite{Rasp} se
permiten o no.

\end{itemize}

\bigskip

\noindent
{\bf 3. CONCLUSI\'ON}

\bigskip

\noindent
En conclusi\'on podemos declarar que presentamos una manera muy natural de
derivar las ecuaciones en la representaciones $(j,0)\oplus
(0,j)$, que lleva a las que fueron dadas por otros
investigadores en el pasado.  Se sabe que la f\'{\i}sica actual del
neutrino ha tropezado en encontrarse con dificultades serias.
Ningun experimento ni observaci\'on est\'a en acuerdo con predicciones
te\'oricas del modelo estandar.  En cambio, nosotros ya ense\~namos de
refs.~\cite{SG,Fush1,Simon} (cf. tambi\'en
art\'{\i}culo reciente~\cite{Rasp}) que
las preguntas relacionadas con toma el l\'{\i}mite correcto sin masa
no son triviales.  Espero que la pregunta de si las ecuaciones propuestas
tienen  pertinencia a la descripci\'on del mundo f\'{\i}sico, se
resolver\'a en un futuro cercano.

\bigskip

{\it Agradecimientos.} El trabajo fue motivado por los art\'{\i}culos
del Prof. D. V. Ahluwalia, por discusiones muy francas y \'utiles,
as\'{\i} como conversaciones telef\'onicas con Prof. A.~F.~Pashkov
durante los \'ultimos 15 a\~nos, por Prof. A. Raspini que bondadosomente
me envi\'o sus recientes art\'{\i}culos, y por el informe cr\'{\i}tico del
\'arbitro an\'onimo de la revista ``Fundaci\'on de F\'{\i}sica" que, no
obstante, estaba de importancia crucial en comenzar mi investigaci\'on
actual.  Tambi\'en, reconozco las discusiones con los Profs. A.  E.
Chubykalo, Y. S.  Kim, R. M.  Santilli y Yu.~F.~Smirnov.  Muchas gracias
a todos ellos.

Reconozco la ayuda en la ortograf\'{\i}a
espa\~nola del Sr.  Cornelio Alvarez.

Zacatecas Universidad, M\'exico, se agradece por la concesi\'on
del profesorado.  Este trabajo ha sido apoyado en parte por
el Sistema Nacional de Investigadores y el Programa de Apoyo
a la Carrera Docente.

\end{document}